\documentstyle[twoside,epsf]{article}

\catcode`\@=11
\long\def\@makefntext#1{
\protect\noindent \hbox to 3.2pt {\hskip-.9pt  
$^{{\eightrm\@thefnmark}}$\hfil}#1\hfill}		%CAN BE USED 

\def\@makefnmark{\hbox to 0pt{$^{\@thefnmark}$\hss}}	%ORIGINAL 
	
\def\ps@myheadings{\let\@mkboth\@gobbletwo
\def\@oddhead{\hbox{}
\rightmark\hfil\eightrm\thepage}   
\def\@oddfoot{}\def\@evenhead{\eightrm\thepage\hfil
\leftmark\hbox{}}\def\@evenfoot{}
\def\sectionmark##1{}\def\subsectionmark##1{}}

\oddsidemargin=\evensidemargin
\newcounter{sectionc}\newcounter{subsectionc}\newcounter{subsubsectionc}
\renewcommand{\section}[1] {\vspace{12pt}\addtocounter{sectionc}{1} 
\setcounter{subsectionc}{0}\setcounter{subsubsectionc}{0}\noindent 
	{\tenbf\thesectionc. #1}\par\vspace{5pt}}
\renewcommand{\subsection}[1] {\vspace{12pt}\addtocounter{subsectionc}{1} 
	\setcounter{subsubsectionc}{0}\noindent 
	{\bf\thesectionc.\thesubsectionc. {\kern1pt \bfit #1}}\par\vspace{5pt}}
\renewcommand{\subsubsection}[1] {\vspace{12pt}\addtocounter{subsubsectionc}{1}
	\noindent{\tenrm\thesectionc.\thesubsectionc.\thesubsubsectionc.
	{\kern1pt \tenit #1}}\par\vspace{5pt}}

\newcommand{\textlineskip}{\baselineskip=13pt}
\newcommand{\smalllineskip}{\baselineskip=10pt}

\def\eightcirc{
\begin{picture}(0,0)
\put(4.4,1.8){\circle{6.5}}
\end{picture}}
\def\eightcopyright{\eightcirc\kern2.7pt\hbox{\eightrm c}} 

\newcommand{\copyrightheading}[1]
	{\vspace*{-2.5cm}\smalllineskip{\flushleft
      {\footnotesize Mod. Phys. Lett. A 18 (2003)   {\bf Brief Review} #1}\\
       {\footnotesize $\eightcopyright$\, World Scientific Publishing Company
        }\\
	 }}

\def\abstracts#1#2#3{{
	\centering{\begin{minipage}{4.5in}\baselineskip=10pt\footnotesize
	\parindent=0pt #1\par 
	\parindent=15pt #2\par
	\parindent=15pt #3
	\end{minipage}}\par}} 

\renewenvironment{thebibliography}[1]
	{\frenchspacing
	 \ninerm\baselineskip=11pt
	 \begin{list}{\arabic{enumi}.}
        {\usecounter{enumi}\setlength{\parsep}{0pt}     
	 \setlength{\leftmargin 12.7pt}{\rightmargin 0pt} %FOR 1--9 ITEMS
         \setlength{\itemsep}{0pt} \settowidth
	{\labelwidth}{#1.}\sloppy}}{\end{list}}

\newcounter{itemlistc}
\newcounter{romanlistc}
\newcounter{alphlistc}
\newcounter{arabiclistc}

\def\@citex[#1]#2{\if@filesw\immediate\write\@auxout
	{\string\citation{#2}}\fi
\def\@citea{}\@cite{\@for\@citeb:=#2\do
	{\@citea\def\@citea{,}\@ifundefined
	{b@\@citeb}{{\bf ?}\@warning
	{Citation `\@citeb' on page \thepage \space undefined}}
	{\csname b@\@citeb\endcsname}}}{#1}}

\newif\if@cghi
\def\cite{\@cghitrue\@ifnextchar [{\@tempswatrue
	\@citex}{\@tempswafalse\@citex[]}}
\def\citelow{\@cghifalse\@ifnextchar [{\@tempswatrue
	\@citex}{\@tempswafalse\@citex[]}}
\def\@cite#1#2{{$\null^{#1}$\if@tempswa\typeout
	{IJCGA warning: optional citation argument 
	ignored: `#2'} \fi}}

\def\@refcitex[#1]#2{\if@filesw\immediate\write\@auxout
	{\string\citation{#2}}\fi
\def\@citea{}\@refcite{\@for\@citeb:=#2\do
	{\@citea\def\@citea{, }\@ifundefined
	{b@\@citeb}{{\bf ?}\@warning
	{Citation `\@citeb' on page \thepage \space undefined}}
	\hbox{\csname b@\@citeb\endcsname}}}{#1}}

\def\@refcite#1#2{{#1\if@tempswa\typeout
        {IJCGA warning: optional citation argument
	ignored: `#2'} \fi}}

\def\refcite{\@ifnextchar[{\@tempswatrue
	\@refcitex}{\@tempswafalse\@refcitex[]}}

\def\pmb#1{\setbox0=\hbox{#1}
	\kern-.025em\copy0\kern-\wd0
	\kern.05em\copy0\kern-\wd0
	\kern-.025em\raise.0433em\box0}

\def\fnt#1#2{\footnotetext{\kern-.3em
	{$^{\mbox{\scriptsize #1}}$}{#2}}}

\def\runninghead#1#2{\pagestyle{myheadings}
\markboth{{\protect\footnotesize\it{\quad #1}}\hfill}
{\hfill{\protect\footnotesize\it{#2\quad}}}}
\font\tenrm=cmr10
\font\tenit=cmti10 
\font\tenbf=cmbx10
\font\bfit=cmbxti10 at 10pt
\font\ninerm=cmr9

\font\eightrm=cmr8

\def\qed{\hbox{${\vcenter{\vbox{			%HOLLOW SQUARE
   \hrule height 0.4pt\hbox{\vrule width 0.4pt height 6pt
   \kern5pt\vrule width 0.4pt}\hrule height 0.4pt}}}$}}

\begin{document}

\newpage

\runninghead{Quantum hamiltonians} 
{prime numbers}

\normalsize\textlineskip
\thispagestyle{empty}
\setcounter{page}{1}

\copyrightheading{}    %{Vol. 0, No.0 (1992) 000--000}

\vspace*{0.88truein}

%\fpage{1} %%%%%%%%%%%%%%%%%%%%%%%%%%%%%%%%%%%%%%%%%%%%%%%%%%%%%%%%%%%
%\centerline{Mod. Phys. Lett. A 16 (2001) xxx
%[gr-qc/0003108 v4]}
\bigskip
\centerline{\bf QUANTUM HAMILTONIANS AND PRIME NUMBERS}  
%\centerline{\bf IN BAROTROPIC OPEN FRW COSMOLOGIES}  
%INDICES\footnote{
%This essay received a ``honorable mention'' in the 
%        A E Competition 
%        F for the year 000 --- Ed.}} 
\vspace*{0.035truein}
%\centerline{\bf MANUSCRIPTS USING COMPUTER SOFTWARE\footnote{For
%the title, try not to use more than 3 lines. Typeset the title
%in 10 pt Times Roman, uppercase and boldface.}}
\vspace*{0.37truein}
%\centerline{\footnotesize NAME}
%\footnote{Typeset names in
%10 pt Times Roman, uppercase. Use the footnote to indicate the
%present or permanent address of the author.}}
%\vspace*{0.015truein}
%\centerline{}
%\baselineskip=10pt
%\centerline{\footnotesize\it City, State ZIP/Zone,
%Country\footnote{State completely without abbreviations, the
%affiliation and mailing address, including country. Typeset in 8
%pt Times Italic.}}
\vspace*{10pt}
\centerline{\footnotesize HARET C. ROSU}
\vspace*{0.015truein}
%\centerline{\footnotesize [Received 17 August 1999] }
\baselineskip=10pt
\centerline{\footnotesize hcr@ipicyt.edu.mx}
\centerline{\footnotesize  Applied Mathematics, IPICyT,} 
\centerline{\footnotesize Apdo Postal 3-74 Tangamanga, 78231 San Luis Potosi,
Mexico
%Instituto de F\'{\i}sica,
%Universidad de Guanajuato, Apdo Postal E-143, 37150 Le\'on, Gto, Mexico
}
\vspace*{0.225truein}
%\publisher{(May 15, 2000)}{}%{(May 14, 2000)}
%\centerline{\footnotesize Received (16 May 2000) }

%%%%%%%%%%%%%%%%%%%%%%%%%%%
\vspace*{0.21truein}
\abstracts{A short review of Schr\"odinger hamiltonians for which the spectral problem  has been related  in the literature to the distribution of the prime numbers is presented here. We notice a possible connection between prime numbers and centrifugal inversions in black holes and suggest that this remarkable link could
be directly studied within trapped Bose-Einstein condensates. In addition, when referring to the factorizing operators of Pitkanen and Castro and collaborators, we 
perform a mathematical extension allowing a more standard supersymmetric approach.
}{}{}
%%%%%%%%%%%%%%%%%%%%%%%%%%%

%\vspace*{10pt}
%\keywords{The contents of the keywords}

\textlineskip                  %) USE THIS MEASUREMENT WHEN THERE IS
\vspace*{12pt}                 %) NO SECTION HEADING

\vspace*{1pt}\textlineskip	%) USE THIS MEASUREMENT WHEN THERE IS
%\section{General Appearance}    %) A SECTION HEADING
\vspace*{-0.5pt}
\noindent

%%%%%%%%%%%%%%%%%%%%%%%%%%%%%%%%%%%%%%%%%%%%%%
PACS number(s):  98.80.Hw, 11.30.Pb

\noindent
%%%%%%%%%%%%%%%%%%%%%%%%%%%%%%%%%%%%%%%%%%%%%%%%%%%%%%%%%%%%%%%%%%%%%

%\newpage

%\pagebreak

%\textheight=7.8truein
%\setcounter{footnote}{0}
%\renewcommand{\thefootnote}{\alph{footnote}}

\bigskip
\bigskip

%\section{The Main Text}

{\bf 1. Introduction}

\medskip

\noindent
The problem of the nontrivial zeros of Riemann's zeta function, i.e., whether they all lie on the line $z=\frac{1}{2}+it$ in the complex plane or not,
is a famous unsolved mathematical problem in which physics, especially quantum mechanics and chaos theory, could have 
a substantial and rewarding contribution in view of its direct connection with the distribution of prime numbers.\cite{mwatkins} 
The phase of Riemann's zeta function could be considered the main mathematical object able to trace this important arithmetic distribution.

In this work, my first goal  is to provide a short survey of the hamiltonians that
so far have been proposed to give hints for a spectral solution (also known as the Hilbert-Polya conjecture) of the location of the zeta  zeros on the critical line.
Along the review the reader can encounter several relevant novel implications in this highly interesting topic, especially in sections {\bf 2} and {\bf 4}. 
%The phase of Riemann's zeta function could be considered the main mathematical object able to trace this important arithmetic distribution. 
Since the logderivative of the phase shift of the Coulomb repulsive potential with reversed centrifugal barrier is a function that approximates very well the Riemann phase for
the negative orbital number $l=-1/4$ and because centrifugal inversions have been conjectured in black holes a promising connection between prime numbers
and black hole solutions in general relativity is suggested. Moreover, this could be under direct experimental investigation in the not quite far future if sonic black hole
configurations could be produced in trapped Bose-Einstein condensates. In addition, when addressing the factorizing operators of Pitkanen and 
Castro and collaborators, a simple mathematical extension of their work is dealt that brings their approach closer to the standard supersymmetric methods of 
quantum mechanics.

\medskip

%%%%%%%%%%%%%%%%%%%%%%%%%%%%%%%%%%%
{\bf 2.  Bhaduri, Khare, Low (1995)} \cite{bkl}

\medskip

\noindent
BKL showed that the density of zeros of Riemann's $\zeta$ function is determined by its phase $\theta(t)$
%----------------------------------
\begin{equation} \label{bkl1}
\exp[2i\theta(t)]=\exp(-it\ln \pi)R_{\Gamma}(t)~, \quad R_{\Gamma}(t)=\frac{\Gamma\left(\frac{1}{4}+\frac{it}{2}\right)}{\Gamma\left(\frac{1}{4}-\frac{it}{2}\right)}~,
\end{equation}
%---------------------------------
where the convention $\theta (0)=\pi$ has been introduced. Although this phase is smooth, i.e., it does not include the jumps by
$\pi$ due to the zeros of the modulus of $\zeta$, it counts well the zeros on the critical line.
It can also be expressed as the logderivative of the $l=-\frac{1}{4}$ phase shift (${\rm arg}\Gamma\left(\frac{3}{4}+\frac{it}{2}\right)$)
with respect to the distorted Coulomb wave of the repulsive Coulomb potential
%---------------------------------
\begin{equation}\label{bkl2}
H_{bkl}=\frac{d^2}{d y^2}-\frac{l(l+1)}{y^2}-\frac{\epsilon}{y}+k^2~,
\end{equation}
%------------------------------
where $l=-\frac{1}{4}$, $\epsilon=-\frac{mE}{2\hbar ^2}$, and $k=\frac{m\omega}{2\hbar}$. %[Bhaduri, Khare, Law].

The potential in Eq.~(\ref{bkl2}) is also
equivalent to an inverted harmonic half-oscillator. Of course, such a negative fractional phase shift does not look physical
at first glance
as it points to an {\em attractive} centrifugal barrier (i.e., well) but this is precisely the case for its application to 
prime numbers. Even more, I hail this as a very appealing feature because of a paper of Cirone {\em et al},\cite{ciro} where
the $-1/4$ correction is called the {\em quantum anti-centrifugal force} being a metric effect related to the radial derivatives
in the Laplacian. Furthermore it is well established that a `centrifugal force reversal' occurs for $r<3m$ in a Schwarzschild 
spacetime.\cite{perlick} The connection between prime numbers and black holes is of course a distinguished one and of
much transcendence. 

It is not hopeless to succeed in touching experimentally the promising connections alluded above. Especially suitable is the BEC physics. According to Garay,
sonic black holes could be
created in BECs trapped by means of sufficiently tight ring-shaped external potentials produced with state-of-the-art or planned technology.\cite{garay} In this case,
the sonic black holes are stationary solutions of the Gross-Pitaevskii nonlinear equation displaying a boundary  from which only a tiny thermal quantum sound radiation 
can escape. Of course, a detailed study of the instablility of these sonic horizons is needed as recently emphasized by Leonhardt and collaborators.\cite{leon}

%-----------------------------------------
%\begin{equation}\label{sbh}
%\Phi(\theta, \tau)=\sqrt{\rho(\theta)}e^{i\int v(\theta)d\theta}~,
%\end{equation}
%-------------------------------------

But most promising, in 2001, Ott {\em et al},\cite{ott} reported the production of a BE condensate in a microstructured magnetic surface trap. This is remarkable because the booming realm of surface-patterning technological processes opens up for the beautiful physical
phase-shift approach to prime numbers. Thus, by means of micro-fabricated electrical circuits, precision measurement
of scattering properties becomes feasible and interferometric measurements should allow soon a direct access of the phase shifts that was never available in 
the past.\cite{brand}  

%%%%%%%%%%%%%%%%%%%%%%%%%%%%%%
\vskip 1ex
\centerline{
\epsfxsize=220pt
\epsfbox{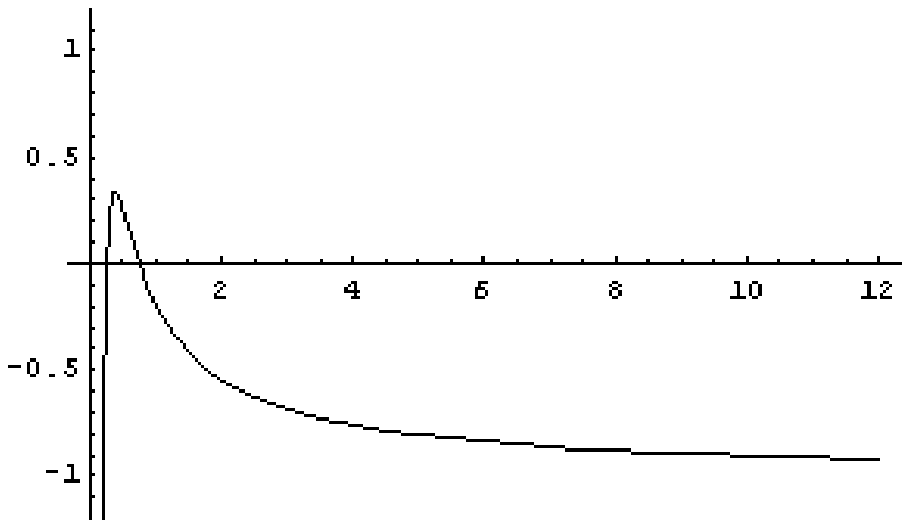}}
\vskip 3ex
\begin{center}
{\small{Fig. 1}$\,$
The BKL Coulomb potential with  $-1/4$ centrifugal correction and the other parameters scaled to unity.}
\end{center}
%%%%%%%%%%%%%%%%%%%%%%%%%%%%%%%%

\bigskip

%%%%%%%%%%%%%%%%%%%%%%%%%%%%%%%%%%%
{\bf 3. Berry and Keating (1999)} \cite{bk}

\medskip

\noindent
This is in the class of Berry and Keating contributions, who were stimulated by ideas and remarks of Connes.
According to Berry and Keating, the hamiltonian 
%------------------------------
\begin{equation}
H_{bk}=-i\hbar\left(x\partial _{x}+\frac{1}{2}\right)
\end{equation}
%----------------------------
is the simplest hermitian operator
connected to Riemann's $\zeta$, though it is only a canonically rotated form of the upturned 
harmonic oscillator $p^2-x^2$. The simplicity results from avoiding the complications of the parabolic cylinder eigenfunctions.
The linear independent wave functions in the $x$ representation are of the type
%---------------------------------------
\begin{equation} \label{bk1}
\psi _{\pm}(x;E)=\frac{A\theta(\pm x)}{|x|^{1/2-iE/\hbar}}~,
\end{equation}
%-----------------------------------------
where $\theta$ is the Heaviside step function. One can also write them  in terms of the smoothed counting function of the Riemann zeros
%%%%%%%%%%%%%%%%%%%%%%%%%%
\begin{equation} \label{bk2}
\psi _{\pm}(x;E)=\frac{\exp[-i\pi(N_{sm}(E)-1]}{|x|^{1/2-iE/\hbar}}~,
\end{equation} 
%%%%%%%%%%%%%%%%%%%%%%%%%%
where Berry and Keating define $N_{{\rm sm}}(E)=\theta_R(E)/\pi+1$, and
%---------------------------
\begin{equation}
\theta _R(E)=-\frac{E}{2}\log \pi + Im\log \Gamma \left(\frac{1}{4}+\frac{1}{2}iE\right)~.
\end{equation}
%------------------------------

In addition, Aneva,\cite{anev} considered chaos quantization conditions for $H_{bk}$ and their geometrical (group) interpretation in an effort to get a discrete 
spectrum for $H_{bk}$.

%%%%%%%%%%%%%%%%%%%%%%%%%%%%%%%%%%%

\bigskip

\newpage

%%%%%%%%%%%%%%%%%%%%%%%%%%%%%%%%%%%%
{\bf 4. Castro et al (2001, 2002)} \cite{castro}

\medskip

\noindent
Castro and collaborators, based on works of Pitkanen,\cite{pit} introduced the simple and appealing nonrelativistic supersymmetric quantum mechanical (SUSYQM) schemes for the Riemann zeros problem. They also elaborated on p-adic oscillators for the same problem. Their first order ``ladder" operators are
\begin{equation}
D_1=-\frac{d}{d\ln t}+\frac{dV(t)}{d\ln t}+k_c \quad {\rm and} \quad D_2=\frac{d}{d\ln t}+\frac{dV(1/t)}{d\ln t}+k_c~, 
\end{equation}
where $k_c$ is a constant parameter and $V(t)$ is given by the logarithm of the Gauss-Jacobi theta series $G(t^l)$, $e^{2V(t)}=G(t^l)$, $l={\rm const}$
\begin{equation}
G(t^l)=\sum _{n=-\infty}^{\infty}e^{-\pi n^2t^l}
\end{equation}
and $V(1/t)\neq V(t)$. The resulting pair of susy partner hamiltonians have the same  {\em real} spectrum $s(1-s)$ on both the real line (trivial zeros) and 
Riemann's critical line (nontrivial zeros). The `coherent' eigenfunctions are written $\Psi _s(t)$ and 
$\Psi _s(1/t)$, respectively, where
\begin{equation}\label{pit}
\Psi _s(t)=t^z\Bigg[\frac{\exp(-t)}{1-\exp(-t)}\Bigg]^{\frac{1}{2}}~.
\end{equation}
The Mellin transform of the Gauss-Jacobi theta series is the zeta function $\zeta(s)$ and the inner product of two eigenfunctions,
say $\Psi _{s1}$ and  $\Psi _{s2}$, can be written in terms of  $\zeta(a(s^{*}_{1}+s_2)+b-\frac{a}{2})$, where $a$ and $b$ can be expressed in terms of 
$k_1$, $k_2$ and $l$. The zeta zeros are in one-to-one correspondence with the `orthogonality' conditions.

\medskip

Here I will show that there is a standard way of doing SUSYQM for this important case. Using a relationship of Gauss that was mentioned by Castro
\begin{equation}
G(1/x)=x^{1/2}G(x)~,
\end{equation}
one can write the above ladder operators in the form
\begin{equation}
D_1=-\frac{d}{d\ln t}+\frac{dV(t)}{d\ln t}+k_1 \quad {\rm and} \quad D_2=\frac{d}{d\ln t}+\frac{dV(t)}{d\ln t}+k_2~, 
\end{equation}
where $k_2=k_1 +\frac{l}{4}$.

%\begin{equation}
%H^{\pm}_{\rm cas}=- \frac{d^2}{(d\ln t)^2} \pm \frac{1}{2}\frac{d^2\ln\Theta (t)}{(d\ln t)^2}+\left(\frac{1}{2}\frac{d \ln \Theta (t)}{d\ln t} + k\right)^2~,
%\end{equation}
%where $\Theta$ is Jacobi's theta series. 

This leads to
\begin{equation}
H_A\Psi _A=D_2D_1\Psi _A=[-d_t^2+(k_1-k_2)d_t +d_tV+V^2+(k_1+k_2)V+k_1k_2]\Psi _A
\end{equation}
and
\begin{equation}
H_B\Psi _B=D_1D_2\Psi _B=[-d_t^2+(k_1-k_2)d_t-d_tV+V^2+(k_1+k_2)V+k_1k_2]\Psi _B~,
\end{equation}
where $d_t=\frac{d}{d\ln t}$ and $V=\frac{1}{2}\ln G(t^l)$.

Using now the gauge transformation $\Psi _{A,B}=\psi _{A,B}t^{l/4}$ one can eliminate the first derivative $d_t$ and get the Schroedinger equations
\begin{equation}
d_t^2 \psi _{A,B}+S_{A,B}\psi _{A,B}=\lambda _{k_1,k_2} \psi _{A,B}~,
\end{equation}
where $\lambda _{k_1,k_2}=(\frac{k_1-k_2}{2})^2-k_1k_2$ plays the role of eigenvalue and $S_{A,B}$ are in the position of Schroedinger susy partner potentials
\begin{equation}
S_{A,B}=\pm d_tV+V^2+(k_1+k_2)V~.
\end{equation}

%On the other hand
%\begin{equation}
%H_B\Psi _B=D_1D_2\Psi _B=[-d_t^2+(k_1-k_2)d_t-d_tU+U^2+(k_1+k_2)U+k_1k_2]\Psi _B~.
%\end{equation}

%%%%%%%%%%%%%%%%%%%%%%%%%%%%%%%%%%%%%%

\bigskip

%%%%%%%%%%%%%%%%%%%%%%%%%%%%%%%%%%%%%%%
{\bf 5. Chadan and Musette (1993)} \cite{cm}

\medskip

\noindent
These authors used
%---------------------------------------------
\begin{equation} \label{cm1}
H _{cm}=-\frac{d^2}{dx^2} -\frac{\Big[l(l+1)-f_{cm}(x;g,\alpha , \beta)\Big]}{x^2}~,
\end{equation}
%----------------------------------------
where the centrifugal correction of Chadan and Musette is
%---------------------------------
\begin{equation}\label{cm2}
F_{cm}=\frac{f_{cm}(x;g,\alpha , \beta)}{x^2}=\frac{\frac{g}{L^2(x)}-\frac{\alpha}{L(x)}-\frac{\beta}{ L(x)LL(x)}}{x^2}~.
\end{equation}
In Eq.~(\ref{cm2})
$L=\log (1/x)$, $LL=\log \log (1/x)$, $l=1$, $\alpha =-27/4$, $\beta =-3/2$ and $g$ is a coupling constant.

\noindent
Chadan and Musette proposed the above  rather complicated singular potential in a closed interval [0,$R$] and Dirichlet boundary conditions at both ends.
They gave arguments that the spectrum in the coupling constant $g=\frac{1}{4}+\frac{t^2}{4}$ ($t$ is the imaginary part on the critical line), 
which is real and discrete, with $g_n>1/4$, coincides approximately with the nontrivial Riemann zeros when $R=e^{-4\pi/3}$.

We note that this is a so-called Sturmian quantum problem, i.e., a quantization problem in the coupling constant of the potential.
A very detailed analysis of this singular hamiltonian and a generalization thereof from the point of view of inverse scattering and $s$-wave Jost functions  
has been performed in the important work of Khuri.\cite{khuri} 

%%%%%%%%%%%%%%%%%%%%%%%%%%%%%%
\vskip 0.05 ex
\centerline{
\epsfxsize=166pt
\epsfbox{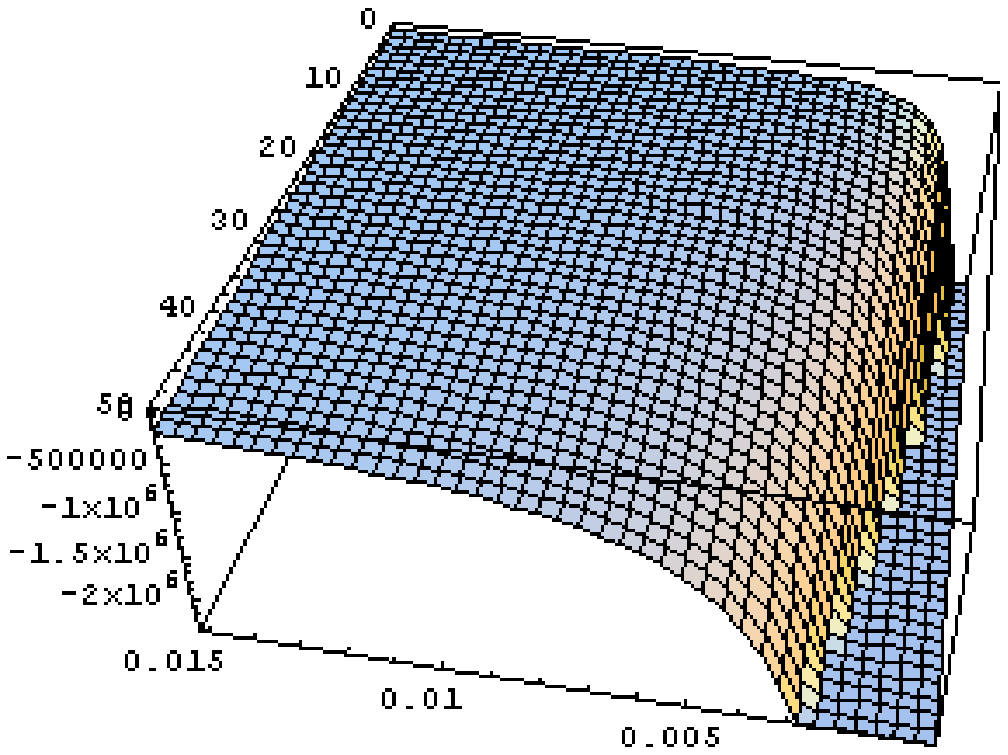}}        %prim2
%\vskip 1ex
\begin{center}
{\small{Fig. 2}$\,$
The centrifugal correction $F_{cm}$ of Chadan and Musette for $R\leq 0.015$ and $g\leq 50$.}
\end{center}
%%%%%%%%%%%%%%%%%%%%%%%%%%%%%%%%

\bigskip

%%%%%%%%%%%%%%%%%%%%%%%%%%
{\bf 6. Joffily (2003): zeros of s-wave Jost functions} \cite{joff}

\medskip

\noindent
The first to relate the nontrivial zeta zeros to the complex poles of a scattering matrix were Pavlov and Fadeev,\cite{pf} for the case of a particle on a 
surface of negative curvature.

Joffily, on the other hand, suggested a `potential scattering' Hilbert-Polya conjecture. He argued that the zeros of Riemann's zeta function could be put in one-to-one correspondence with the zeros of the s-wave Jost function for cutoffed potentials in the complex 
momenta plane. Joffily considered the s-wave Jost function $f_{+}(\gamma)$ in the dimensionless argument $\gamma =kR$ ($R$ is the cutoff of the potential) 
and the transformation
%----------------------------------------
\begin{equation} \label{jof1}
z=-i\frac{\gamma}{2Im \gamma}
\end{equation}
%------------------------------------------
by which the lower half of complex $\gamma$-plane is mapped onto the critical zeta line.

%%%%%%%%%%%%%%%%%%%%%%%%%%%%%%%%%%%%%%%

\bigskip

%%%%%%%%%%%%%%%%%%%%%%%%%%%%%%%%%%%%%%%%
{\bf 7.  The fractal potential of Wu and Sprung (1993)} \cite{ws}

\medskip

\noindent
Wu and Sprung obtained a local potential given by an Abel integral equation based on the smooth number of Riemann zeros below $E$, $N(E)=\frac{E}{2\pi}\ln \frac{E}{2\pi e}+\frac{7}{8}$, considered as the semiclassical WKB density of states below $E$ of a potential.
Their potential $y(x)=\frac{V}{V_0}$ scaled with respect to $V(0)=V_0$ (Wu and Sprung chose $V_0=3.10073\pi$) is given by
\begin{equation}
%x=\frac{1}{\pi}\Bigg[ \sqrt{V-V_0}\ln \frac{V_0}{2\pi e^2}+\sqrt{V}\ln \frac{\sqrt{V}+\sqrt{V-V_0}}{\sqrt{V}-\sqrt{V-V_0}}\Bigg]
x=\frac{\sqrt{V_0}}{\pi}\Bigg[ \sqrt{y-1}\ln \frac{V_0}{2\pi e^2}+\sqrt{y}\ln \frac{\sqrt{y}+\sqrt{y-1}}{\sqrt{y}-\sqrt{y-1}}\Bigg]~.
\end{equation}
\noindent
As a function of the fitted Riemann zeros, $y(x;N)$ shows fluctuations with respect to the smooth potential $y(x)$ and therefore $y(x;N)$ is a fractal curve. Using 
a box-counting-method analysis,
Wu and Sprung obtained the result that the first 500 Riemann nontrivial zeros can be fitted with the  fractal potential $y(x;500)$ of fractal dimension 1.5.

\bigskip

%%%%%%%%%%%%%%%%%%%%%%%%%%%%%%%%%%%%%%%
{\bf 8. Okubo's two-dimensional potential (1998)} \cite{ok}

\medskip

\noindent
Okubo introduced
%--------------------------------
\begin{equation}
H_{\rm ok}(x,y;\beta _{\rm ok})=\frac{\partial ^2}{\partial x \partial y}+i\beta  _{\rm ok}y \frac{\partial}{\partial y}+i(1-\beta  _{\rm ok})x\frac{\partial}{\partial x}+\frac{i}{2}~,
\end{equation}
%-----------------------------
where $\beta  _{\rm ok}$ is a real parameter.
He gave arguments that this two-dimensional Lorentz-invariant Hamiltonian may be relevant to the RH. 
Some eigenfunctions of his two-dimensional Hamiltonian corresponding to infinite-dimensional representation of the Lorentz group have many interesting properties. 
Especially, a relationship exists between the zero zeta function condition (ZZFC) that should be satisfied by any nontrivial zeta zero
%----------------------------------
\begin{equation}\label{zzfc}
\int _0^{\infty}d\tau \,\frac{\tau^{z-1}}{1+\exp \tau}=0~,
\end{equation}
%-----------------------------------
for $Re\,z \in (0,1)$ and the absence of trivial representations in the wave function.
The eigenfunctions of $H_{\rm ok}\phi=\lambda \phi$ with $z=\frac{1}{2}+i\lambda$ are of the form
%-----------------------------------
\begin{equation} \label{phiok}
\phi(x,y)=\int _0^{\infty}d\tau \,\tau^{z-1}\exp (ix\tau ^{1-\beta  _{\rm ok}})g(\tau+y\tau^{\beta  _{\rm ok}})~,
\end{equation}
%-----------------------------------
where $g$ is an arbitrary function which vanishes as its argument goes to $\infty$. The choice
%----------------------
\begin{equation} \label{g0}
g_0(\tau+y\tau ^{\beta  _{\rm ok}})=\frac{1}{1+\exp(\tau +y\tau ^{\beta  _{\rm ok}})}
\end{equation}
%---------------------
yields a function $\phi _0$ which is an infinite-dimensional {\em unitary} realization of the Lorentz group SO(1,1). If the so-called ZZFC condition is applied to
$\phi _0$ the representation space of SO(1,1) does not contain any singlet representation of the group.

Berry and Keating commented that Okubo's hamiltonian is a more general, relativistic oscillator than their own quantized $xp$.

%%%%%%%%%%%%%%%%%%%%%%%%%%%%%%%%%%%%%%%

\bigskip

{\bf 9. Mussardo's potential (1997) }\cite{mus}

\medskip

\noindent
Mussardo's potential $W$ is given by

\begin{equation}
x(W)=\sum _{m=1}^{\infty} \frac{\mu(m)}{m}\int _{E_0}^{W}\frac{\epsilon ^{\frac{1-m}{m}}}{\ln \epsilon \sqrt{W-\epsilon}}\,d\epsilon~,
\end{equation}
where $\mu$ is the M\"obius function.

\noindent
Since Mussardo applies, like Wu and Sprung, the semiclassical method, his work is based on Abel's integral equation. All that we said with respect
to the work of Wu and Sprung applies here as well.

Mussardo also suggested a resonance experiment to implement the primality test up to a maximum number $N$ depending on a cutoff $\epsilon _0$. Sharp resonances in the transmission amplitude $T_N(E)$ of plane 
waves impinging on the potential $W(x;\epsilon _0(N))$ would indicate that the numbers $N$ are prime.

%%%%%%%%%%%%%%%%%%%%%%%%%%%%%%%%%%%%%%%

\bigskip

{\bf 10. de Oliveira and Pellegrino (2001)} \cite{op}

\medskip

\noindent
These authors considered discrete Schr\"odinger operators of the type
\begin{equation} \label{op1}
(H_Vu)_n=u_{n+1}+u_{n-1}+\lambda V_nu_n
\end{equation}
defining $V_0=0$, $V_n=1$ if $n+1$ is a prime number, and $V_n=0$ if not. They have found a localization-delocalization transition as a function of the 
potential strength $\lambda$.

%%%%%%%%%%%%%%%%%%%%%%%%%%%%%%%%%%%%%%%%%%%%%%%5

\bigskip

{\bf 11. Many-body hamiltonians: Boos and Korepin (2001, 2002)} \cite{kor}

\medskip

\noindent
Many-body hamiltonians are interesting more from the statistical physics point of view than the spectral one. Boos and Korepin,\cite{kor} showed that 
particular correlation functions in quantum spin chains can be expressed in terms of the values of Riemann zeta function at odd arguments (for even arguments,
it is known that $\zeta (2n)\sim B_{2n}$, the even Bernoulli numbers). They put forward 
the following conjecture

\medskip

\noindent
{\em Arbitrary correlators of the $XXX$ antiferromagnet are described as certain combinations of $\ln 2$, the Riemann zeta function with odd arguments and 
rational coefficients}.

\medskip

An extension of this result to complex arguments could be interpreted as decay (decoherence) of the spin chain correlators.   

\medskip

We recall that the $XXX$ finite $N$ chain antiferromagnet is described since 1928 by the following Heisenberg hamiltonian
\begin{equation} \label{bk1}
H_{XXX}=\sum _{i=1}^{N}(\sigma _i^x\sigma _{i+1}^{x}+\sigma _i^y\sigma _{i+1}^y+\sigma _i^z\sigma _{i+1}^z-1)~,
\end{equation}
where the $\sigma$'s are Pauli matrices.  

\bigskip

%%%%%%%%%%%%%%%%%%%%%%%%%%%%%%%%%%%%%%%%%%%%%%%%%%%%%%%%%
{\bf 12. Crehan (1995) } \cite{cre}

\medskip

\noindent
Crehan used a classical theorem of Hardy and Littlewood stating that the position of the $n$th prime on the critical line is limited by $t_n <a^{-1}n$
($a={\rm const}$) to get a theorem asserting that there is an infinite family of classically integrable nonlinear oscillators whose quantum spectrum is given by the 
imaginary part of the sequence of zeros on the critical line of the Riemann zeta function. In addition, Forrester and Odlyzko,\cite{fo} related the distribution of 
zeta critical zeros to generalized Painleve fifth differential equation.

\medskip

%\newpage

%%%%%%%%%%%%%%%%%%%%%%%%%%%%%%%%%%%%%%%%%%%%%%%%%

{\bf 13. Conclusion}

\medskip

\noindent
The spectral interpretation of the imaginary parts of the nontrivial Riemann zeros stimulated mathematical physicists to propose several quantum 
hamiltonians with spectra that could be useful in tackling with this century-old problem. Here the suggestions made in this area
are gathered together in order to catch a global glimpse of these research facts. I provide a number of useful comments that hopefully will be taken into
account in future possible experiments. The most exciting issue is the possibility of direct experimental access to negative phase shifts in the physics of trapped BECs.
This cutting-edge research is of direct relevance for the link between prime numbers and black holes. I also hint that the rapidly developing {\bf PT} quantum 
mechanics,\cite{bender} a specific type of non-hermitic hamiltonians symmetric with respect to parity and time reversal discrete operations that
display real spectra, will have an important contribution in this field. In fact, the susy hamiltonian pair of Pitkanen and Castro and collaborators are not hermitic but have real spectra.

%%%%%%%%%%%%%%%%%%%%%%%%%%%%%%%%%%%%%%%%%%%%%%%%%%%%%%%%%%%%%%%%%%%%%%%%

\bigskip

{\bf Acknowledgements}

\medskip

\noindent
I wish to thank the Editor for useful remarks that led to the improvement of presentation and V. Korepin, M. Novello and C. Castro who concerned to email me 
their comments on the arXiv preprint of this work (quant-ph/0304139).

\bigskip

%\newpage

%\bibitem{das} A. Das, {\em Integrable Models}, World Scientific, 1989;
%For recent work, see e.g., A. Das and Z. Popowicz, Phys. Lett. A {\bf 274},
%30 (2000) and references therein.

%\newpage

%%%%%%%%%%%%%%%%%
%\vskip 2ex
%\centerline{
%\epsfxsize=180pt
%\epsfbox{hubble1.eps}}
%\vskip 12ex
%\begin{center}
%{\small{Fig. 1}:$\quad$%\\
%The function $dH_{11}^{0}/dt+
%(H_{11}^{0})^{2}=\frac{d^2a_{11}/dt^2}{a_{11}}$ 
%showing that for this
%superfield component there is 
%cosmological acceleration (of different strength) for all values of $c$
%(here $c\in [-0.1, 1]$, $t\in[0.15, 2.5])$. At small times,
%one can see a deflationary phase.}
%\end{center}
%%%%%%%%%%%%%%%%

%%%%%%%%%%%%%%%%%
%\vskip 2ex
%\centerline{
%\epsfxsize=180pt
%\epsfbox{hubble2.eps}}
%\vskip 12ex
%\begin{center}
%{\small{Fig. 2}:$\quad$%\\
%The same plot as in Fig. 1 for the same time interval and 
%$c\in [-0.9, -0.1]$ . One can see the strong 
%acceleration effect corresponding to fluids in the neighborhood of the 
%cosmological vacuum and also traces of the deflationary phase.}
%\end{center}
%%%%%%%%%%%%%%%%

\end{document}